# Oxygen Content Variation and Cation Doping Dependence of $(La)_{1.4}(Sr_{1-y}Ca_y)_{1.6}Mn_2O_{7\pm\delta}$ ($y = 0, 0.25, 0.5$) Bilayered Manganites Properties

Lorenzo Malavasi [a,*], Maria Cristina Mozzati [b], Cristina Tealdi [a], Carlo B. Azzoni [b], and Giorgio Flor [a]

[a] Dipartimento di Chimica Fisica "M. Rolla", INSTM, IENI/CNR Unità di Pavia of Università di Pavia, V.le Taramelli 16, I-27100, Pavia, Italy.

[b] CNISM, Unità di Pavia and Dipartimento di Fisica "A. Volta", Università di Pavia, Via Bassi 6, I-27100, Pavia, Italy.

## Abstract

The results of the synthesis and characterization of the *optimally* doped $(La)_{1.4}(Sr_{1-y}Ca_y)_{1.6}Mn_2O_{7\pm\delta}$ solid solution with $y=0$, 0.25 and 0.5 are reported. By progressively replacing the Sr with the smaller Ca, while keeping fixed the hole-concentration due to the divalent dopant, the "size effect" of the cation itself on the structural, transport and magnetic properties of the bilayered manganite has been analysed. Two different annealing treatments of the solid solution, in pure oxygen and in pure argon, allowed also to study the effect of the oxygen content variation. Structure and electronic properties of the samples have been investigated by means of X-ray powder diffraction and X-ray absorption spectroscopy measurements. Magnetoresistivity and static magnetization measurements have been carried out to complete the samples characterization. Oxygen annealing of the solid solution, that showed a limit for $y\sim0.5$, induces an increase of the Mn average valence state and a transition of the crystal structure from tetragonal to orthorhombic while the argon annealing induces an oxygen under-stoichiometry and, in turn, a reduction of the Mn average valence state. Along with the Ca substitution, the Jahn-Teller distortion of the $MnO_6$ octahedra is reduced. This has been directly connected to a general enhancement of the transport properties induced by the Ca-doping. For the same cation composition, oxygen over-stoichiometry leads to higher metal-insulator transition temperatures and lower resistivity values. Curie temperatures ($T_C$) reduce by increasing the Ca-doping. The lower $T_C$ for all the annealed samples with respect to the "as prepared" ones are connected to the strong influence on the magnetic interaction of the point defects due to the $\delta$-variation.

*Corresponding Author: Dr. Lorenzo Malavasi, Dipartimento di Chimica Fisica "M. Rolla", INSTM, Università di Pavia, V.le Taramelli 16, I-27100, Pavia, Italy. Tel: +39-(0)382-987921 - Fax: +39-(0)382-987575 - E-mail: lorenzo.malavasi@unipv.it

# 1. Introduction

Perovskite manganites represent nowadays one of the most intensively studied research topic in the fields of solid state chemistry and physics. This is due to several reasons among which the high negative magnetoresistive effect, termed colossal (CMR), and the virtual infinite tunability of the physical properties of the $R_{1-x}A_xMnO_3$ perovskite structure by means of cation doping and/or oxygen content control [1-8].

More recently, the discovery of the CMR effect in the $La_{2-2x}Sr_{1+2x}Mn_2O_7$ manganite, which represents the *n*=2 member of the Ruddlesden-Popper series of manganites, has attracted special interest due to its crystal structure forming a naturally layered system [9-13]. In particular, distinct features of the bilayered manganites are the anisotropic characteristics in both charge-transport and magnetic properties and the reduced dimensionality of the Mn–O–Mn networks which leads to several intriguing changes including enhanced MR effects, large magneto-caloric effects, unconventional magnetostriction, and anisotropic transport in charge carriers. Moreover, these layered systems exhibit a variety of both ferromagnetic and antiferromagnetic structures.

The generic formula of the R-P phases is $A_{n+1}Mn_nO_{3n+1}$ which corresponds to the staggering of *n* perovskites layers intercalated by a rock-salt block. For *n*=∞ the perovskites manganites are obtained while for *n*=1 the antiferromagnetic (AF) or spin-glass insulator phases $La_{1-x}Sr_{1+x}MnO_4$ are known [14,15].

In the *n*=2 member $La_{2-2x}Sr_{1+2x}Mn_2O_7$ an extremely rich variety of magnetic phases have been found as a function of the Sr-doping. One of the most interesting regions extend from *x*=0.3 to *x*=0.4 where ferromagnetic metals (FMM) are found. However, also within a so relatively narrow doping range several different kinds of arrangements of manganese ions spin occur. A common properties is the presence of a 2D ferromagnetic ordering in each perovskite layer and between the $MnO_2$ layers even though with different spin directions as a function of *x*. At *x*=0.3, the magnetic moments of each $MnO_2$ layer couple ferromagnetically within a bilayer and antiferromagnetically, along the *c*-axis, between successive bilayers. At $x \approx 0.32$ the inter-bilayer coupling becomes FM but still directed along the *c*-axis. At $x \geq 0.33$ the magnetic moments direct along the *ab*-plane. The magnetic coupling between the constituents single $MnO_2$ layers changes from FM into canted-AF beyond $x \sim 0.4$ [15]. Also the magnetic coupling above the transition temperature ($T_C$) is rather interesting. As suggested by several groups [16-17] a 2D ferromagnetic coupling within the plane should occur and the presence of a FM-AFM correlation between each plane in the bilayer unit, for $x=0.4$, has been pointed out. In any case, there exists two-dimensional ferromagnetic short-range order in a wide temperature region above $T_C$ which may be related to the anisotropic exchange energy ($|J_{ab}| > |J_c| >> |J'|$, $J_{ab}$, $J_c$, and $J'$ standing for the in-plane, inter-single-layer, and inter-bilayer exchange interaction, respectively) in the quasi-two-dimensional FM system.

As in the perovskite manganites, a close coupling between the magnetic and transport properties is observed. However, due to the strong anisotropy that characterizes these layered systems, peculiar features have been observed. Always considering the $0.3 \leq x \leq 0.4$ range for La$_{2-2x}$Sr$_{1+2x}$Mn$_2$O$_7$, single-crystals studies have shown that the ratio between the resistivity along the *c*-axis ($\rho_c$) and the resistivity in the *ab*-plane ($\rho_{ab}$) is as large as $10^2$ at room temperature (RT), which suggests a confinement of the carrier motion within the $MnO_2$ bilayer. Usually, the on-set of the long-range FM order is accompanied by a resistivity drop (the insulator-to-metal transition, IM) followed by a metallic-like transport. In the *T*-range above $T_C$ both $\rho_c$ and $\rho_{ab}$ show an activated-like transport for $x=0.4$ with hopping energies around 30-40 meV [18]. By lowering the doping the nature of the insulating state in the $\rho_{ab}$ is progressively suppressed and, at $x=0.3$, a metal-like transport ($d\rho_{ab}/dT > 0$) is observed in the range $T_C \leq T \leq 270$ K [9].

The doping-level dependence of the crystal structure of the bilayered manganites has been object of several studies [19-21]. All the members of the La$_{2-2x}$Sr$_{1+2x}$Mn$_2$O$_7$ solid solution, for $0.3 \leq x \leq 0.5$, possess, at RT, a tetragonal structure belonging to the space group *I4/mmm*.

In particular, it has been shown that the *a(b)* lattice parameter slightly increases with the increase of *x*, whereas the *c* parameter decreases more rapidly [21]. The lattice parameters variation reflects changes in the bond lengths. Taking into account the relative trend of the three distinct Mn-O bond lengths *vs*. *x* it turned out that the collective J-T distortion increases by decreasing the *x*-value. Since the reduction of the cation doping corresponds to $e_g$-electron doping, the systematic change of the Mn-O bond lengths reflects the electron occupancy of the two $e_g$-orbital states [21]. The significant expansion of the out-of-plane Mn-O(2) bond (the one not directly linked to the other perovskite layer) by lowering the Sr-doping is consistent with a shift of electron density from the planar ($x^2$-$y^2$) to the axial ($3z^2$-$r^2$) orbitals.

The role of A-site doping on the bilayered manganites properties has been studied as done in the case of the perovskites manganites. However, fewer dopants have been considered and for less extensive doping levels due not only to the fact that the field of bilayered manganites is relatively new with respect to that of the perovskites, but also due to less straightforward preparation routes.

Battle *et al*. [22] considered the effect of La replacement with several other lanthanides (keeping Sr as the divalent dopant) and concluded that the size of the cation strongly influences the stability of the R-P phase and the A cations distribution between the two available crystallographic sites. Smaller lanthanides showed a preference in occupying the rock-salt layer with respect to the perovskite block. Moreover, they pointed out that cation disorder in compounds with larger lanthanides is at the origin of a subtle phase separation into two *n*=2 R-P phases [22].

Calcium doping on the A-site has been object of an early investigation by Asano *et al*. who studied the La$_{2-2x}$Ca$_{1+2x}$Mn$_2$O$_7$ solid solution for $0 \leq x \leq 0.5$ [23]. The main conclusions of that work is the clear evidence of two distinct type of FM ordering which possibly results from anisotropic exchange interactions. Moreover, the material, in the doping region $0.22 \leq x \leq 0.5$, undergoes two transitions from a paramagnetic-insulator to a ferromagnetic-insulator and finally to a ferromagnetic- metal with decreasing temperature. A model considering the *intra*- and *inter*-layers spin arrangement was proposed.

Taking into account the available literature regarding the synthesis and characterization of bilayered manganites it appears that further work is needed. A fundamental issue which has been well recognized in the perovskites manganites is the role and control of oxygen stoichiometries on their physical and chemical properties. We have clearly showed that any variation in the oxygen stoichiometry ($\delta$) with respect to the ideal value of 3 not only induces a change of the Mn valence state but strongly influences the manganites properties through the point defects connected to the $\delta$-variation, namely cation vacancies ($\delta > 0$) or anion vacancies ($\delta < 0$) [7,8,24,25]. We remark here that the presence of oxygen over-stoichiometry ($\delta > 0$) is not properly accounted for in the formalism $LaMnO_{3+\delta}$ since it is well established that the extra-oxygen is not of interstitial-type but rather cation vacancies are formed for compensation [25-27]. So, the correct formula for the oxygen over-stoichiometric perovskites should be: $La_{1-\varepsilon}Mn_{1-\varepsilon}O_3$, where $\varepsilon = \delta/(3+\delta)$.

The role of oxygen content variation has never been systematically considered in any layered manganite (also encompassing the $n=1$ member of the R-P family). Only one thorough study by Milburn and Mitchell has recognized the importance of this topic in the $Nd_{2-2x}Sr_{1+2x}Mn_2O_7$ system ($0.10 \leq x \leq 0.70$), where they analyzed the phase diagram as a function of composition and oxygen partial pressure [28]. The Authors also considered the distribution of oxygen vacancies and found a close relationship with the composition and Mn average oxidation state. However, the characterization of the samples was mainly structural and so a direct link with the physical properties was not given.

In this paper we present the result of the synthesis and characterization of the $(La)_{1.4}(Sr_{1-y}Ca_y)_{1.6}Mn_2O_{7\pm\delta}$ solid solution for $y=0$, 0.25, 0.5. We chose this composition since it represents an *optimally* doping analogous to the $x=0.3$ doping in the $La_{1-x}A^{2+}_xMnO_3$ perovskite manganites. In addition, keeping fixed the hole-concentration due to the divalent cation, we changed its size by progressively replacing the Sr with the smaller Ca. The extent of this solid solution showed a limit for $y \sim 0.5$ [29]. Finally, in order to vary the oxygen content, the samples underwent two different annealing treatments, *i.e.* in pure oxygen and pure argon.

Samples so prepared have been fully characterized for what concerns their structure, by means of X-ray powder diffraction (XRPD), and their electronic properties, by means of X-ray absorption spectroscopy (XAS) measurements, which were mainly used to determine the Mn valence state. Magnetoresistivity (MR) and static magnetization ($M$) measurements have been carried out to complete the samples characterization.

Aim of this paper is to show the effect of the oxygen content variation and a possible "size effect" of the divalent A-cation on the structural, transport and magnetic properties of an optimally doped bilayered manganite.

# 2. Experimental

(La)$_{1.4}$(Sr$_{1-y}$Ca$_y$)$_{1.6}$Mn$_2$O$_{7\pm\delta}$ samples with $y$=0, 0.25 and 0.50 were synthesized by solid state reaction starting from proper amounts of La$_2$O$_3$, Mn$_2$O$_3$, SrCO$_3$ and CaCO$_3$ (Aldrich >99.99%). Pellets were prepared from the thoroughly mixed powders and allowed to react first at 1273 K for 72 hours and after at 1573 K for other 72 hours. During these thermal treatments the pellets were re-ground and re-pelletized at least three times. The as prepared samples have been annealed in an home made quartz apparatus at 1173 K and at $P$(O$_2$) = 10$^5$ Pa (pure oxygen) and 10$^{-1}$ Pa (argon) for 48 hours and cooled down to RT at 20 K/min.

XRPD patterns were acquired on a Bruker "D8 Advance" diffractometer equipped with a Cu anode, graphite monochromator on the diffracted beam and proportional detector. Measurements were carried out in the angular range from 10° to 120° with a step size of 0.02° and a counting time of 10 s per step. Diffraction patterns were refined by means of Rietveld method with the FULLPROF software [30].

Magnetisation measurements were carried out with a SQUID magnetometer by applying different magnetic fields (0 - 7 T) in the temperature range 2 - 320 K. MR measurements were carried out between 320 and 10 K at various fields with the DC-four electrodes method by means of a specific probe directly inside the SQUID apparatus.

Mn-K edges XAS spectra were collected in transmission mode at RT at the BM-08 beam line (GILDA) of the ESRF synchrotron radiation laboratory (Grenoble, France) using ion chambers as detectors. For all the measurements, the samples were mixed with cellulose and pressed into pellets. The amount of sample in the pellets was adjusted to ensure a total absorption ($\mu$) above the edge around 2. Spectra were processed by subtracting the smooth pre-edge background fitted with a straight line. Each spectrum was then normalized to unit absorption at 1000 eV above the edge, where the EXAFS oscillations were not visible any more. Spectra processing was done with the ATHENA [31] software. In all the measurements the spectrum of a metallic Mn foil was collected together with the spectra of the samples. This was done in order to give reliable data about the edge position.

# 3. Results and Discussion

## 3.1 X-ray powder diffraction

Figure 1 reports the indexed diffraction pattern of the $La_{1.4}Sr_{1.6}Mn_2O_7$ manganite after the synthesis and prior to any thermal treatment (called in the following "as-prepared" (AP) samples) while a graphical representation of this structure is depicted in Figure 2. The pattern can be perfectly indexed according to a tetragonal unit cell belonging to the *I4/mmm* space group (no. 139). Lattice constants for this sample are $a=b=3.8682(2)$ Å, $c=20.274(1)$ Å and cell volume 303.359(3) Å$^3$. These values are also reported in Table 1 together with the lattice constants for the as-prepared $La_{1.4}Sr_{1.2}Ca_{0.4}Mn_2O_7$ and $La_{1.4}Sr_{0.8}Ca_{0.8}Mn_2O_7$ samples which were indexed in the *I/4mmm* space group, as well. The lattice constant trend shows a progressive and significant reduction of the *c* parameter as the Ca-doping increases while the *a(b)* parameter first decreases passing to the $La_{1.4}Sr_{1.2}Ca_{0.4}Mn_2O_7$ composition and then increases when the Ca/Sr ration is equal to 1. However, an overall reduction of the cell is found, as shown by the cell volume behaviour. This is in agreement with the ionic radii difference between the $Sr^{2+}$ and $Ca^{2+}$ for the same coordination number, *i.e.* 1.40 and 1.26 Å, respectively.

Let us now pass to consider the effect of thermal annealing treatments on the samples. Figure 3 compares the XRPD patterns around the main peak for the $La_{1.4}Sr_{1.6}Mn_2O_7$ sample, chosen as a representative example, annealed in pure oxygen ("O-annealed" in the Figure) and pure argon ("Ar-annealed" in the Figure) with the as-prepared sample. From the patterns for the Ca undoped sample it is clear that after the oxygen annealing a relevant broadening of the main peak located around 32° is observed. It passes from a FWHM ~0.125° for the as-prepared sample to ~0.240° after the oxygen annealing, while the next peak, at about 32.5°, undergoes a reduced broadening from ~0.127° to ~0.143°. The first peak corresponds, in the tetragonal *I4/mmm* structure, to the (105) diffraction line while the second one to the (110). The presence of this broadening, also found for other peaks in the pattern, may be indicative of a phase transition towards a less symmetric structure. As reported by Milburn and Mitchell [28] two distortions of the $I4/mmm$ cell to lower (orthorhombic) symmetry are known. The first one consists of a doubling of the cell size in a ($\sqrt{2}a$) $\times$ ($\sqrt{2}b$) $\times$ $c$ superstructure, with a rotation of 45° around the $c$ parameter, in which the face diagonals of the $I4/mmm$ cell become the new cell vectors and the $\langle100\rangle$ mirror planes are lost. When this distortion occurs, the tetragonal $\{hhl\}$ peaks split into $\{2h,0,l\}$ and $\{0,2h,l\}$, the tetragonal $\{hkl\}$ into $\{h+k,h-k,l\}$ and $\{h-k,h+k,l\}$ while the $\{h0l\}$ reflections remain unsplit. The second type of distortion involves the loss of the $\langle110\rangle$ mirror planes and gives origin to an orthorhombic cell of the same volume as the tetragonal one in which the $a$ and $b$ cell directions are no longer equivalent in magnitude. Such distortion has been reported for mixed valence compounds such as the $K_2NiF_4$-type nickelates [32-33] and for the $n = 2$ layered manganites of the $Nd_{2-2x}Sr_{1+2x}Mn_2O_7$ system [28] where, however, the orthorhombic transition was found for oxygen deficient samples ($\delta<0$) and not, as in the present case, for oxygen over-stoichiometric ones ($\delta>0$). For this second distortion, the $\{hhl\}$ reflections should remain degenerate whereas the $\{h0l\}$ split into $\{h0l\}$ and $\{0hl\}$. This structure is described in the $Immm$ space group (no. 71).

By looking carefully through the XRPD patterns of the oxygen annealed samples we could observe the broadening of all the $\{h0l\}$ reflections, thus confirming the occurrence of the second type of distortion. In fact, as shown by Figure 3, the peak that corresponds to the single (105) reflection in the as-prepared sample, belonging to the $I4/mmm$ space group, has been split into two unresolved diffraction lines in the $Immm$ symmetry, *i.e.* the (105) and (015) reflections. However, beside this evidence, some peaks belonging to the $\{hhl\}$ type showed significant broadening in particular for the $\{0010\}$ and $\{1110\}$ reflections which should not be affected by the $I4/mmm \rightarrow Immm$ symmetry change. This is shown in detail in the inset of Figure 3, which presents the $2\theta$ regions where the $\{0010\}$ and $\{1110\}$ reflections are located, for the $La_{1.4}Sr_{1.6}Mn_2O_7$ as-prepared, oxygen-annealed and argon-annealed samples.

However, this broadening effect has not the same extension for all the cation stoichiometries considered here. This is clearly shown in Figure 4 where the same three regions of Figure 3 are plotted for the $La_{1.4}Sr_{1.6}Mn_2O_7$, $La_{1.4}Sr_{1.2}Ca_{0.4}Mn_2O_7$ and $La_{1.4}Sr_{0.8}Ca_{0.8}Mn_2O_7$ samples annealed in pure oxygen. By increasing the Ca-doping the broadening of *all* the peaks, both belonging to the *Immm* symmetry and the other ones are less broadened by the oxygen thermal treatment. For example, the FWHM for the as-prepared and oxygen-annealed $La_{1.4}Sr_{0.8}Ca_{0.8}Mn_2O_7$ sample is practically constant.

Let us consider now the strategy employed to treat the patterns of the samples annealed in pure oxygen. Regarding the Ca-undoped compound, a refinement with a single *Immm* phase did not lead to a good fit result ($R_B = 12.1$, $\chi^2 = 5.04$). Several models were considered in order to improve the refinement. It was found that a better description of the experimental pattern could be achieved by means of a multi-phase model. Also in this context different combinations of tetragonal and orthorhombic phases were considered. An improvement of the data fit was obtained by means of a two *Immm*-phases model ($\chi^2 = 3.66$). This allowed a better description of most of the peaks but the ones interested by significant broadening as the {0010} were not yet properly refined.

The results for $La_{1.4}Sr_{1.6}Mn_2O_7$ indicate the presence of about 62% of an orthorhombic phase with lattice constants $a=3.8688(5)$, $b=3.8683(5)$ and $c=20.175(3)$ and a second phase (about 38%) with lattice constants $a=3.8666(6)$, $b=3.8698(7)$ and $c=20.016(2)$. As can be noticed, the orthorhombic distortion is small for each phase and the main difference between them resides in the value of the parameter *c*. Probably, this difference arises from a difference in the relative occupancies of the A cations of the two available sites, *i.e.* in the perovskite or in the rock salt layer, and also/or from an extremely small compositional difference between the two phases that could be related to a variation of the mean oxidation state of Mn ions.

Also for the $La_{1.4}Sr_{1.2}Ca_{0.4}Mn_2O_7$ sample the agreement between the experimental and calculated patterns was achieved with a two-phase model which resulted in a better fit with respect to the previous sample ($\chi^2 = 2.45$) and gave 60% of a phase with $a=3.8598(4)$, $b=3.8610(4)$ and $c$=19.880(2) and about 40% of a second phase with $a$=3.8618(5), $b$=3.8601(4) and $c$=20.011(2). For this sample and for the previous one both sets of lattice constants are reported in Table 1.

Finally, for the sample with an equal Sr/Ca ratio, the pattern refinement was reliable starting with a single phase model. However, also in this case, the best agreement between the experimental and calculated patterns was achieved with an orthorhombic structure ($R_B = 13.2$, $\chi^2 = 3.6$) instead of a tetragonal one ($R_B = 19.6$, $\chi^2 = 5.7$). Lattice constants for this sample are listed in Table 1, as well.

When the as-prepared samples have been annealed in pure argon the tetragonal structure has been preserved and no evidence of an anomalous broadening of any peak was detected. Again, in Table 1, are reported all the lattice parameters for the argon-annealed $La_{1.4}Sr_{1.6}Mn_2O_7$, $La_{1.4}Sr_{1.2}Ca_{0.4}Mn_2O_7$ and $La_{1.4}Sr_{0.8}Ca_{0.8}Mn_2O_7$ samples.

Figure 5 summarizes the lattice constants variation by plotting the trend of cell volumes against the Ca content ($y$), thus referring to the formula $(La)_{1.4}(Sr_{1-y}Ca_y)_{1.6}Mn_2O_{7\pm\delta}$. Data are relative to the three cation compositions and to the as-prepared (white circles), oxidised (black triangles) and reduced (black squares) samples. According to the refinement results presented above, for the oxidised $La_{1.4}Sr_{1.6}Mn_2O_7$ and $La_{1.4}Sr_{1.2}Ca_{0.4}Mn_2O_7$, two values of the volume are reported in the Figure.

For the same cation composition the annealing treatment in pure oxygen induces a reduction of the cell volume with respect to the as-prepared samples. This links directly to the increase of the average oxidation state of Mn ions (being the ionic radii 0.645 Å and 0.53 Å for $Mn^{3+}$ and $Mn^{4+}$, respectively [34]) as confirmed by the XAS results (see later). In addition, for $y$=0 and 0.25 this kind of treatment induces a phase separation in the samples giving origin to two distinct orthorhombic phases with similar $a$ and $b$ constants (differences lower than 0.1%) and more different $c$ parameters ($\Delta c$~0.8-0.6%).

As suggested before, this phase separation is probably due to the effect of oxygen content increase. In 3D perovskites such as $LaMnO_3$ it is well known that for $\delta$>0 the point defects compensations occurs via cation vacancies formation according to [26]:

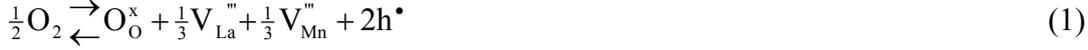

$$\frac{1}{2}O_2 \underset{\leftarrow}{\rightarrow} O_O^x + \frac{1}{3}V_{La}''' + \frac{1}{3}V_{Mn}'''' + 2h^\bullet \qquad (1)$$

The defect chemistry of 2D layered manganites has not been object of analogous studies apart for the (Nd,Sr)MnO system in presence of oxygen under-stoichiometry [28]. For oxygen over-stoichiometric samples we can just postulate the occurrence of a mechanism similar to that reported by equation 1 by considering, at a first approximation, that the perovskite layer, which is structurally analogous to the 3D perovskites, should behave in the similar way. In addition, also for the rock-salt layer we can suppose that the oxygen over-stoichiometry will be compensated by cation vacancies rather than oxygen interstitials as found in binary oxides with this crystal structure [35].

The formation of cation vacancies may be at the base of the phase separation as a consequence of difference in their distribution and/or clustering phenomena. In particular, the actual distribution of cation vacancies may be not trivial since they can form on both the independent crystallographic sites corresponding to cations in the perovskite block, 2b (00½), and in the rock-salt layer, 4e (00$z$). In addition, vacancies may prefer to localize on each site as a preference to create unequal numbers of cation vacancies, *i.e.*, for example, again referring to eq. 1, a ratio of $V_{La}''' / V_{Mn}'''' \neq 1$, was found to occur in 3D perovskites [25].

Opposite to the oxygen annealing, the treatment in pure argon, at fixed cation composition, induces an increase of the cell volumes for all the samples, as shown in Figure 5. We remark here that the Vegards law is observed for all the monophasic sample series and that the solubility limit can be ruled out as the source of the aforementioned phase separation. Regarding the solubility limit of Ca in the $La_{1.4}Sr_{1.6}Mn_2O_7$ compound we confirm the available literature results [29] which indicate a threshold around 0.5. In fact, any attempt to synthesize composition with a Ca/Sr ratio greater than 1 were unsuccessful. In all the cases, a perovskite phase together with SrO oxide were found to be the stable reaction products.

The cell volume increase found for the reduced samples can be connected to the reduction of the Mn sub-lattice as a consequence of the reduction of the oxygen content. This is in accordance with the XAS data (see later) which pointed out a Mn-edge shift towards lower energies for the argon-annealed samples with respect to the samples treated in pure oxygen.

Looking at the lattice constant trend of the argon-annealed samples it is interesting to note an anisotropy in the lattice parameters variation, *i.e.* the progressive reduction of the *c*-parameter and the increase of the in-plane parameters which results in a progressive reduction of the *c/a* ratio. This result may be indicative of a variation in the Jahn-Teller distortion of the $MnO_6$ octahedra triggered by a variation in the nature of the orbital state of the $e_g$ electrons where, for high Ca-contents levels, the electrons prefer to occupy the $3x^2$-$y^2$ orbitals. This can due to the effect of Ca-doping on the electronic structure of the bilayered manganites. Even though the atomic position determined from the Rietveld refinement may not be fully reliable, particularly with reference to the oxygen atoms, the values of the Mn-O bond lengths have been used to determine the distortion of the octahedron by means of the Jahn-Teller parameter, defined as [13]:

$$Q = 4\{(l+m)/2 - s\} \qquad (2)$$

where *l*, *m* and *s* represent the long, medium and short Mn-O bonds in the $MnO_6$ octahedron. By means of this expression, it is found that the *Q* value moves from 0.2952 to 0.2482 and finally to 0.1526 by increasing the Ca-content. This result witnesses a reduction of the J-T distortion by increasing the Ca-doping while keeping the overall hole content approximately constant.

## 3.2 X-ray Absorption Spectroscopy

XAS measurements at the Mn-K edge were carried out for the $La_{1.4}Sr_{1.6}Mn_2O_7$, $La_{1.4}Sr_{1.2}Ca_{0.4}Mn_2O_7$ and $La_{1.4}Sr_{0.8}Ca_{0.8}Mn_2O_7$ samples annealed in oxygen and argon. As a typical example, Figure 6 reports the edge region spectra of the $La_{1.4}Sr_{1.6}Mn_2O_7$ sample annealed in pure oxygen ($La_{1.4}Sr_{1.6}Mn_2O_7$-O) and in pure argon ($La_{1.4}Sr_{1.6}Mn_2O_7$-Ar) together with the spectra of the two standards employed in this study: $LaMnO_3$ ($Mn^{3+}$ standard) and $CaMnO_3$ ($Mn^{4+}$ standard). A general view of the spectra of all the samples is presented in Figure 7 with a vertical reference line passing through the maximum of the first sample.

The spectral line shapes and widths recorded are very similar for all the samples. A chemical shift of the threshold energies ($E_0$) (taken at the inflection point of the absorption edge) towards higher energies is found for the samples annealed in oxygen with respect to those annealed in argon, as shown clearly in Figure 6 for $La_{1.4}Sr_{1.6}Mn_2O_7$. This chemical shift can be related to the increase of the average Mn valence which is connected to the creation of holes in the rather localized $d$ orbitals with consequent loss of screening.

The overall shift from +3 ($LaMnO_3$) to +4 ($CaMnO_3$) Mn valence was around 3.3 eV, in good agreement with previous available literature data [36]. Through the chemical shifts ($\Delta E_0$), measured with respect to the $LaMnO_3$ threshold, and considering a roughly linear behaviour of the edge position with the average Mn oxidation state [36-37] and our previous XAS data on 3D manganites [38], we estimated the mean Mn valence state for each sample. We note that for the samples annealed in pure oxygen the Mn valence state is the same for the $y$=0.25 and 0.50 compositions ($\sim 3.38$) and slightly lower for the Ca-undoped sample ($\sim$3.34). A significant reduction of the Mn valence state occurs after the argon annealing with estimated values of 3.27, 3.25 and 3.27 for the $y$=0, 0.25 and 0.5 samples, respectively. If one consider that an acceptable error on the valence evaluation through this procedure may be around 3-5%, also taking into account the data resolution, we note that the valence state within each of the two sample series may be thought to be nearly constant. This consideration will allow us to discern the influence of the divalent dopant on the physical properties of the samples considered.

Based on these results, the oxygen annealed samples possess an oxygen over-stoichiometry ($\delta$) of the order of 0.06 while those annealed in argon have an oxygen under-stoichiometry of about 0.04. To the best of our knowledge, this is the first work considering the role of oxygen stoichiometry fluctuations as the main variable to change the average Mn oxidation state of a bilayered manganites, where, usually, the hole doping is realized through the variation of the divalent dopant concentration [10-13-36].

Some additional features of the XAS spectra can be noticed in the pre-edge region. The inset of Figure 6 brings into prominence this part for the samples considered in that Figure and indicates three features, labelled as $A_1$, $A_2$ and $A_3$, on the $CaMnO_3$ spectrum, due to their particular enhancement in this compound. These "peaks" have been correlated to the electronic and structural properties of the 3D manganites [39,37]. They originate from transitions to empty states with $d$-like character ($1s \rightarrow 3d^{(n+1)}$). This kind of transition would be dipole forbidden, however, through an admixture of 3$d$ and 4$p$ states it becomes weakly allowed. In addition, in the pre-edge region there is also some hybridization with the O 2p states. However, these considerations are not enough to account for the high intensity of these $A_i$ features in samples such as the $CaMnO_3$ which presents Mn atoms at a point of inversion symmetry and nearly so even for the substituted $LaMnO_3$. A clarification of this point arose from the calculation of Elfimov [40] who showed that the 3$d$ features correspond to a coupling of Mn 4$p$ states with 3$d$ states on adjacent Mn atoms. A more detailed description of these last aspects can be found in References 36 and 39. Taking into account the calculation reported in the cited papers it has been shown that the $A_1$ peak would correspond to transitions into empty majority-spin $e_g$ states on the neighbouring Mn ions and $A_2$ peak to transitions into the $e_g$ and $t_{2g}$ minority states. A more recent paper suggests [41], as a further assumption, that the 1s to $t_{2g}$ transition intensity vanishes as confirmed by theoretical calculations [42,43].

Figure 8a and 8b put in prominence the pre-edge region for the oxygen and argon annealed samples, respectively, by displaying the spectra in the range 6538-6548 eV. A deep analysis of these features goes beyond the scope of the present paper and will be object of future work together with the analysis of the EXAFS signal; however some general consideration of relevance for the following discussion may be done. First of all the features which are found in the in the pre-edge region of the $n=\infty$ members of the R-P series (*i.e.* the 3D manganites), and which have been discussed above, are also observed for the $n=2$ members. In particular, just two $A_i$ peaks (labelled $A_1$ and $A_2$) may be found in the spectra of our samples with a mean separation of about 2.1-2.2 eV in all the cases. The presence of just two peaks is in accordance with the observation that, in general, the $A_3$ peak is not observed in doped-manganites also for high resolution data. No significant differences in the $A_1$-$A_2$ separation is found between the oxygen and argon annealed samples. Again, this is in agreement with the fact that a total decrease of the splitting of only ~0.3 eV is found when changing the average Mn valence from +3 to +4 [36] and so the relatively small valence change between our samples should not strongly affect this aspect. Within each series, the position of the $A_1$ peak remains practically constant while some shift occurs for the $A_2$ peak. In particular, for the oxygen annealed compounds its position moves toward higher energy for the intermediate composition ($La_{1.4}Sr_{1.2}Ca_{0.4}Mn_2O_7$) and then shifts slightly back to lower energy for the Ca : Sr ratio equal to 1. Also for the reduced samples the $La_{1.4}Sr_{1.2}Ca_{0.4}Mn_2O_7$ compound shows the $A_2$ peak at higher energy with respect to the other two stoichiometries.

Moreover, a significant difference in the relative intensity of the $A_1$ and $A_2$ peaks between samples annealed in different environment occurs; for the same composition, the oxygen treatment increases the intensity of both features. An increase of the $A_1$ peak intensity has been connected to an increase in the charge delocalization of the majority spin $e_g^1$ electron [44] and also to an enhancement of the covalent character which can be connected to the increase in the oxygen content and in turn of the Mn valence state, which should broaden the conduction bandwidth.

Considering the two series separately, it can be noticed that there is a pronounced increase of the intensity of the $A_2$ feature as the Ca-doping increases. As remembered above, this last peak should originate from transitions into $e_g$ and (less) $t_{2g}$ minority states. However, the reason of this enhancement may not be limited to this aspect. In fact, the $CaMnO_3$ (see inset of Figure 6), differently from the pure $LaMnO_3$, shows a particularly intense $A_2$ peak. It has been pointed out by previous workers that the actually available models developed for describing the pre-edge features of manganites lack of some "ingredients" such as the i) local distortions of the crystal and ii) further change in covalency and hybridization. Since the average Mn valence state is approximately constant within each series of samples it is clear that the enhancement of the $A_2$ feature must have another origin with respect to a simple change in the valence state. The main difference between the $LaMnO_3$ and $CaMnO_3$ spectra is the oxidation state change of Mn from +3 to +4; however, another significant difference is the absence of the Jahn-Teller distortion for the $CaMnO_3$ sample. What we propose is that the intensity of the $A_2$ peak is not only controlled by the average valence state but also by the J-T effect, *i.e.* by local distortions in the crystal structure. As a matter of fact, Figure 9 presents the trend of bond lengths towards the Ca-content for the argon-annealed samples (for the oxygen annealed ones the presence of two phases makes a schematization of the bond lengths trend not so meaningful). As can be appreciated, there is a progressive reduction of the difference between the Mn-O1 and Mn-O3 bonds as a function of $y$, which suggests a reduction of the J-T effect as the Ca-doping proceeds; this is also coupled to a general reduction of the distance between successive perovskites layers. However, we remark that the effect of this last evidence on the physical properties (such as the $T_C$) may be in competition with a reduction of the bandwidth as the mean ionic radius on the A-site reduces and also with a reduction of the statistical variance in the distribution of ionic radii ($\sigma^2$).

*3.3 Electrical Conductivity and Magnetoresistivity*

Electrical conductivity measurements have been carried out on all the annealed samples at null applied magnetic field ($H$) and with a field of 1 and 7 T.

Figure 10 and 11 show the resistivity ($\rho$) curves for the $La_{1.4}Sr_{1.6}Mn_2O_7$ (a), $La_{1.4}Sr_{1.2}Ca_{0.4}Mn_2O_7$ (b) and $La_{1.4}Sr_{0.8}Ca_{0.8}Mn_2O_7$ (c) samples after they have undergone the argon (Fig. 10) and oxygen (Fig. 11) annealing treatments. Inset in each graph presents the magnetoresistivity (MR) defined as:

$$MR(\%) = \frac{\rho(H) - \rho(0)}{\rho(0)} \bullet 100 \ . \tag{3}$$

In all of the $\rho$ *vs*. $T$ curves a transition from an insulating towards a metallic-like behaviour is found. The values of these transitions temperatures, taken at the maximum of the $H$=0 curves and hereafter called $T_\rho$, are reported in Table 2.

For the oxygen annealed samples the $T_\rho$ values progressively reduce by increasing the Ca-doping; this follows the trend reported in literature for the as-prepared samples with analogous compositions even though with higher absolute values [29, 45]. A more complex behaviour is observed for the argon annealed compounds, for which the $T_\rho$ values, always lower with respect to $T_\rho$ values of the as-prepared samples, reduce from 83 to 71 K going from the Ca-undoped sample to the $y$=0.25 one and increase to 100 K for the $y$=0.50 composition.

The absolute $\rho$ values progressively reduce by increasing the Ca-doping for both the types of annealing. This could be directly linked to the reduction of Mn-O bond lengths as the Ca-doping increases, thus making the polaron hopping easier. It is interesting to note that for the argon annealed samples $\rho$ varies in a very large range: the $\rho_{300K}$ reduces of nearly four orders of magnitude, passing from 927 $\Omega$cm for $y$=0 to 2.2 $\Omega$cm for $y$=0.25 and 0.9 $\Omega$cm for $y$=0.50 while for the oxygen annealed samples the $\rho$ values, always lower with respect to the corresponding reduced samples, decrease of about one order of magnitude in the same composition range. Moreover, the width of the transition is relatively large which may be a direct consequence of the point defects connected to the oxygen non-stoichiometry, analogously to what found in perovskite manganites [7,8].

An up-turn of resistivity at low temperature is present for all the three argon annealed samples; this is nearly $T$-independent falling, irrespective to the cation doping, around 32-34 K. On the contrary, for the oxygen ones, the up-turn of the resistivity at low temperature is progressively shifted from ~65 K in $La_{1.4}Sr_{1.6}Mn_2O_7$ to ~40 K in $La_{1.4}Sr_{1.2}Ca_{0.4}Mn_2O_7$ and disappears for the $La_{1.4}Sr_{0.8}Ca_{0.8}Mn_2O_7$ compound.

Coming to the MR behaviour for the argon annealed samples, we note that $La_{1.4}Sr_{1.6}Mn_2O_7$ shows a peaked MR curve with a value, at 7 T, of about 60% and centred at a temperature (~98 K) close to the $T_\rho$ value. As usual for analogous systems, the MR peak moves to higher $T$ as the field is increased. For $La_{1.4}Sr_{1.2}Ca_{0.4}Mn_2O_7$ the MR curve at 1 T has a peaked form with $T_{peak}$ at about 63 K corresponding to a value of ~80% while a higher magnetic field induces a sort of *plateau* behaviour with MR values ranging between 95 and 99.5% for $T$<70 K. Finally, the MR curves for $La_{1.4}Sr_{0.8}Ca_{0.8}Mn_2O_7$ show a nearly linear decrease of MR with temperature at 1 T and a peaked curve at 7 T with $T_{peak}\approx112$ K and a MR effect of about 54%.

For the oxygen annealed series it can be seen that for the Ca-undoped and $y$=0.25 samples just a continuous reduction of MR with $T$ is found while the appearance of a peak followed by an up-turn towards more positive MR values is evident for the $La_{1.4}Sr_{0.8}Ca_{0.8}Mn_2O_7$ sample for all the applied fields (included $H$=0.1T). Absolute values of the MR effect are very close for the $y$=0 and 0.25 compositions and larger for the 0.50 sample, reaching, at 7 T, a maximum value around 60% at ~75 K.

To discuss in more detail the resistivity data we tried to model the activated part of the $\rho$ *vs.* $T$ curves with the usually and widely accepted models, *i.e.* small polaron hopping (adiabatic and non-adiabatic) and variable range hopping (VRH), in order to extract information on the hopping energy.

For all the argon-annealed samples a good fit in the range $1.33T_\rho \leq T \leq 320K$ was achieved considering a small polaron hopping conductivity:

$$\rho = \rho_0 T \exp(E_a / k_B T).$$ (4)

The hopping energies obtained are 128 meV for $La_{1.4}Sr_{1.6}Mn_2O_7$, 75.5 meV for $La_{1.4}Sr_{1.2}Ca_{0.4}Mn_2O_7$ and 52 meV for $La_{1.4}Sr_{0.8}Ca_{0.8}Mn_2O_7$, thus indicating that carrier motion becomes progressively easier as the Ca-doping increases. This is also in agreement with the reduction of resistivity values observed with the increase of calcium doping.

Since the hole doping for the three samples annealed in argon is about the same (see the XAS section for details) the reason for the easier polaron hopping could be connected to the reduction of the Jahn-Teller distortion as the Ca-doping increases, as reported in Section 3.1. We point out that the sample with the highest Ca-substitution for Sr, for which a very low $c/a$ value is detected (5.164(3), see table 1), also displays the highest $T_\rho$ value among the argon annealed series.

For all the three samples annealed in pure oxygen, in a $T$ range far from $T_\rho$, a simple power-law dependence with $T$ was found to describe the transport behaviour in the insulating phase, instead of the "usual" exponential $T$ dependence. For the $y=0$ and $y=0.25$ compositions this could be connected to the presence of a second phase but, interestingly, the single-phase $y=0.50$ composition behaves in the same way, thus suggesting that it may be a property related to the nature of the samples themselves. A possible explanation of this result may come from the known anisotropy in the transport properties between the $a$-$b$ plane and the $c$-axis found in these materials. As reported by Kimura *et al.* [9] for the $La_{1.4}Sr_{1.6}Mn_2O_7$ the conductivity within the layers has a metallic-like trend for temperatures lower than ~270 K while it is always insulating-like between adjacent layers. However, for polycrystalline samples the resistivity behaviour for $T>T_\rho$ is always activated in nature. This suggests that the electrical properties are in some way influenced by the less mobile carrier transport in the out-of-plane direction [23]. So, an enhancement of the carrier motion in this direction, as presumably happens for the oxygen annealed samples, would reduce the strong resistive contribution of the $c$-axis hopping and lead to an overall $\rho$ vs. $T$ trend which may not be modelled with just an activated law.

Differences in the $T_\rho$ between the oxidised and reduced samples can be connected to the difference in the Mn average valence state: higher values for the oxygen-annealed samples (see the 3.2 Section) leads to higher values of the transition temperatures and the lowest $T_\rho$ value is detected for the argon annealed $y$=0.25 composition, for which also the lowest Mn valence state is evaluated.

The different transport behaviour between the two differently annealed series is also strongly related to structural effects. All the three samples annealed in pure oxygen, characterized by $\rho$ values lower and $T_\rho$ values higher with respect to the argon annealed series, have a smaller unit cell with respect to those treated in argon, thanks to the higher oxidation of the Mn ions, and a $c/a$ parameter that, on average, is lower than 5.2: the high $c/a$ values of the argon annealed $y$=0 (c/a $\cong$ 5.27) and $y$=0.25 (c/a > 5.22) compositions indeed correspond to markedly higher $\rho$ values with respect to the other samples (see $\rho_{300K}$ values in Table 2).In addition, we recall the Mn-O bond lengths behaviour we have shown in Figure 9, which points out a progressive reduction of the J-T distortion as the Ca-doping increases. As a consequence, the structural parameters suggest that inter-layers correlations may be enhanced for all the oxidised samples. This could be also an additional reason for the different $T_\rho$ vs. $y$ behaviour for the reduced samples where, at $y$=0.50, an increase of $T_\rho$ is found. Actually, among the argon-annealed samples, only the $y$=0.50 composition has the $c/a$ parameter lower than 5.2.

The overall MR behaviour for the considered samples is the one expected for the CMR manganites. An additional remark concerning the MR effect for $y$=0 and 0.25 oxygen treated samples has to be made. The absence of peaked MR curves in both 1 and 7 T is indicative that other mechanisms in these samples may be predominant with respect to the CMR one. A smooth increase of MR effect with temperature decrease is usually connected to a tunnelling magnetoresistive effect (TMR) which is enhanced as spin scattering is progressively reduced by lowering $T$. This is the common mechanism encountered in multilayers of FM/I/FM alternated layers (I=insulating phase; FM=ferromagnetic phase) and nanosized samples. However, from electronic microscopy the average grain size of the present samples resulted to be in the microns range, so that this origin for the observed MR effect can be ruled out. More probably, for the $y$=0 and 0.25 compositions the reasons of this behaviour may lay in the biphasic nature of the two samples themselves possibly related to the difference in the distribution of cation vacancies and/or clustering phenomena.

## 3.4 Magnetisation Measurements

Figure 12 and 13 report the $M$ vs. $T$ curves at 100 Oe for the three samples annealed in pure oxygen and argon, respectively.

The curves are characterized by a progressive enhancement of the magnetization as the temperature is reduced. In nearly all the curves a further step rise of $M/H$ is found at low $T$. The temperatures of these transitions ($T_C$) are reported in Table 2. It can be noticed that the Ca-doping ($y$) induces, for both the treatments, a progressive reduction of the transition temperatures as $y$ is increased. In addition, as the Ca-doping increases, also the $M/H$ step-rise at low temperature becomes relatively smaller with respect to the $M/H$ slow enhancement from high $T$. This is true for both the under- and over-stoichiometric samples where, for the latter series, at $y$=0.50, the step-rise at low-$T$ has finally disappeared.

The role of cation doping on the magnetic properties of the samples may be understood within the framework of the following model according, also, to what proposed by Asano [23] and observed by Kimura [9]. At room temperature the system in mainly composed of a paramagnetic matrix where some 2D FM in-plane correlations are already present and evolve as $T$ is reduced towards $T_C$. In this region, we recall, the conductivity within the $a$-$b$ planes is already metallic in nature whereas is insulating-like along the *c*-axis. Due to the strong FM coupling of Mn moments starting from ~270 K and to the strong in-plane exchange interaction, the $x$=0.3 ($La_{2-2x}Sr_{1+2x}Mn_2O_7$) compound is considered as a quasi-two-dimensional ferromagnet (2D-FM). Together with the increase of in-plane magnetic correlations also the out-of-plane correlation strengthens with the reduction of *T*, *i.e.* inter-plane FM correlations within constituent $MnO_2$ bilayers. At a defined temperature, usually connected to the drop of $\rho_c$, a 3D FM ordering, where the spin of neighbouring $MnO_2$ layers are aligned along a common direction, takes place. The easy axis of the 3D spin alignment is doping-dependent [10]. Finally, further reducing the temperature from $T_C$ will allow the development of a magnetic order among $MnO_2$ *bi*layers. We also note that the 3D FM order has been observed for an hole doping (*x*) around ~0.30-0.40, where fall our oxygen-annealed samples; in contrast, for the argon-annealed samples, we may not exclude the evolution of AF ordering among $MnO_2$ bilayers for $T < T_C$ [10]. A confirmation of this comes from the observation that, moving from $T_C$ towards lower *T*-values, where the magnetic interaction between $MnO_2$ *bi*layers is progressively developed, by increasing the field, the molar magnetization for the Ar-annealed sample is reduced, thus suggesting that an AF interaction between $MnO_2$ *bi*layers is found, in accordance with the phase diagram developed for $La_{2-2x}Sr_{1+2x}Mn_2O_7$ when $x \leq 0.3$.

The reduction of 3D ordering temperature as Ca-doping increases is connected to the progressive strengthening of the in-plane exchange interaction, $J_{ab}$, with respect to the out-of-plane one, $J_c$. This is in turn linked to the change of the orbital character of the $e_g$ electrons. In detail, as already pointed out in the Results section devoted to the X-ray diffraction analysis, the lattice constants (particularly $c/a$) and bond-lengths trend indicated a progressive shift of the $e_g$ electron density from the $3z^2$-$r^2$ to the $3x^2$-$y^2$ orbitals as the Ca-doping increases; accordingly, it will become *more difficult* to set up a 3D FM ordering among $MnO_2$ layers and $MnO_2$ *bi*layers. However, the transfer integral of the in-plane interaction will be increased by the electron density shift thus making the in-plane carrier hopping easier in accordance with the progressive reduction of the electrical resistivity and hopping activation energy.

This picture is able to account for the decrease of $T_C$ with the Ca-doping, but does not explain the low $T_C$ values determined which, usually, for this doping regime, are of the order of 120-130 K (we underline that the our *as-prepared* La$_{1.4}$Sr$_{1.6}$Mn$_2$O$_7$ sample has a $T_C$ of 115 K).

However, up to now we are not considering the fact that all samples possess an oxygen stoichiometry fluctuation. As we have already put in prominence for the perovskite manganites [7,8,24,25] the point defects connected to the $\delta$-variations are deeply effective in changing the physical properties of manganites. It is clear that even an increase of the Mn valence state, as induced by the oxygen annealing, does not lead to an increase of the magnetic transition temperatures. This has to be directly related to the cation vacancies introduced in the structure along with the oxygen over-stoichiometry. Analogously, the argon annealing treatment does not only induce a reduction of the Mn valence state but also introduce oxygen vacancies in the structure, according to:

$$O_O^x \Leftrightarrow V_O^{\bullet\bullet} + 2e^- + \frac{1}{2}O_2 \qquad\qquad (5)$$

We stress that the above presented picture, in which we suggest a progressive reduction of the 3D FM ordering as the Ca-doping increase, has been recently confirmed by neutron diffraction measurements [46].

Finally, a comment about the significant difference of the $T_C$ compared to the transition temperatures in the resistivity curves. In principle, as the FM order shifts to lower $T$ with Ca-increase, we should expect also worsening effects on the transport data. But this is not the case. The only evident worsening effect caused by the Ca-doping increase is a slight reduction of the $T_\rho$ for the oxygen over-stoichiometric samples. It is clear, from the experimental evidences, that the drop in the resistivity curves, for both oxygen under- and over-stoichiometric samples, is not accompanied by a 3D-FM order within planes, which, in contrast, is set up only for lower $T$. This might suggest that, in the present case, the transition in the resistivity is dominated by the more mobile in-plane carriers and that only for very low temperatures also the contribution from inter-layer transport influences significantly the transport properties. This is also confirmed by the decrease of the $\rho$ up-turn at low-$T$ as $y$ increases, that is when $\rho_c$ decreases as well.

# 4. Conclusion

In the following we summarize the main conclusions we collected from this experimental investigation of the oxygen content variation and cation doping dependence of the $(La)_{1.4}(Sr_{1-y}Ca_y)_{1.6}Mn_2O_{7\pm\delta}$ ($y$=0, 0.25, 0.50) bilayered manganites properties:

1. The solubility limit of the $(La)_{1.4}(Sr_{1-y}Ca_y)_{1.6}Mn_2O_7$ solid solution is around $y$=0.50; further Ca-doping leads to the stabilization of the perovskite phase.

2. Oxygen annealing induces an increase of the Mn average valence state and a transition of the crystal structure from tetragonal to orthorhombic; in addition, for $y$=0 and 0.25, a phase separation in two orthorhombic phases with similar in-planes lattice parameters ($a$ and $b$) and different $c$ parameters is found.

3. Thermal treatments carried out in an argon environment induce an oxygen under-stoichiometry in the samples; this leads, in turn, to a reduction of the Mn average valence state.

4. The Ca-doping manifests in an anisotropic behaviour of the lattice parameters in the tetragonal samples: this is indicative of a change in the electronic occupancy of the $e_g$ electrons as the doping proceeds. In addition, the Jahn-Teller distortion is reduced along with the Ca substitution.

5. XAS analysis was carried out on all the samples and, to the best of our knowledge, it is the first time this kind of study is realised on the $(La)_{1.4}(Sr_{1-y}Ca_y)_{1.6}Mn_2O_7$ solid solution. A deep analysis of the pre-edge features allowed us to propose possible correlations between the pre-edge peaks and the cation doping in the solid solution.

6. Transport properties show a general enhancement induced by the Ca-doping, as witnessed by the resistivity reduction for both the annealed series and lowering of the $E_a$ and the IM transition temperature trend in the reduced samples. This has been directly connected to the reduction of the J-T effect promoted by the calcium doping. For the same cation composition, oxygen over-stoichiometry leads to higher IM transition temperatures and lower $\rho$-values, most probably as a consequence of the higher Mn valence state in the oxidised samples.

7. Ca-doping seems to significantly depress the FM interaction between Mn ions while improving the carrier motion; this may suggest that a strong FM coupling, as expected in the case of a double-exchange mediated transport, is not required for the hole hopping in these layered samples.

8. Curie temperatures reduce by increasing the Ca-doping as a consequence of the progressive strengthening of the in-plane exchange interaction, $J_{ab}$, with respect to the out-of-plane one, $J_c$. This is in turn linked to the change of the orbital character of the $e_g$ electrons.

9. $T_C$ for all the annealed samples are lower with respect to the "as prepared" ones even for the oxidised samples where the Mn valence state is higher; this can be connected to the strong influence on the magnetic interactions of the point defects due to the $\delta$-variation.

10. We finally stress that a significant difference is encountered between the $T_C$ and $T_\rho$ for the same sample; this sort of *uncoupling* between the two transition temperatures seems to be a direct effect of the oxygen non-stoichiometry since the same as-prepared samples presented here behaves in the "usual" way, displaying $T_C$ and $T_\rho$ very close each others. We may propose that this effect is caused by the location of cation and oxygen vacancies introduced in the structure by the oxygen non-stoichiometry; we plan to carry out high resolution neutron diffraction measurements to try to locate these defects and correlate them to the observed effect.

# Acknowledgements

Financial support from the Italian Ministry of University (MIUR) through the PRIN-2004 project is gratefully acknowledged. LM is grateful to the "Accademia Nazionale dei Lincei" for financial support. Dr. Elena di Tullio is acknowledged for sample preparation. ESRF beam line BM08 (GILDA) staff is acknowledged for its support during measurements collection.

# Figures Captions

**Figure 1** – X-ray diffraction pattern of the as-prepared (AP) $La_{1.4}Sr_{1.6}Mn_2O_7$ manganite. Bragg peaks appear as vertical grey lines.

**Figure 2** – Sketch of the $I4/mmm$ crystal structure of the $La_{1.4}Sr_{1.6}Mn_2O_7$ manganite.

**Figure 3** – XRPD patterns around the main peak for the AP, oxygen annealed and argon annealed $La_{1.4}Sr_{1.6}Mn_2O_7$ sample. The inset shows, for the same patterns, the $2\theta$ regions where the $\{0010\}$ and $\{1110\}$ reflections are located.

**Figure 4** – XRPD patterns for the (A) $La_{1.4}Sr_{1.6}Mn_2O_7$, (B) $La_{1.4}Sr_{1.2}Ca_{0.4}Mn_2O_7$ and (C) $La_{1.4}Sr_{0.8}Ca_{0.8}Mn_2O_7$ samples annealed in pure oxygen (the same three $2\theta$ regions of Figure 3).

**Figure 5** – Cell volumes against the Ca content ($y$), for the three cation compositions of the as-prepared (white circles), oxidised (black triangles) and reduced (black squares) samples.

**Figure 6** – Edge region of the XAS spectra of the $La_{1.4}Sr_{1.6}Mn_2O_7$ sample annealed in pure oxygen (blue line) and in pure argon (green line), together with the spectra of the two standards: $LaMnO_3$ (red line) and $CaMnO_3$ (black line). The inset shows an enlarged view of the low-energy part of the same spectra, evidencing the three features labelled as $A_1$, $A_2$ and $A_3$ on the $CaMnO_3$ spectrum.

**Figure 7** – General view of the XAS spectra of all the argon and oxygen annealed samples. The vertical line passing through the maximum of the oxygen annealed $La_{1.4}Sr_{1.6}Mn_2O_7$ sample is drawn as reference.

**Figure 8** – Pre-edge region of the XAS spectra for the (a) oxygen and (b) argon annealed samples.

**Figure 9** – Trend of the three Mn-O bond lengths towards the Ca-content ($y$) for the argon annealed samples.

**Figure 10** – Resistivity curves *vs. T* at 0, 1 and 7 T for the argon annealed (a) $La_{1.4}Sr_{1.6}Mn_2O_7$, (b) $La_{1.4}Sr_{1.2}Ca_{0.4}Mn_2O_7$ (in this case the 7T $\rho$ *vs. T* curve is reported in the inset) and (c) $La_{1.4}Sr_{0.8}Ca_{0.8}Mn_2O_7$. In the insets: correspondent MR curves for the same samples.

**Figure 11** – Resistivity curves *vs. T* at 0, 1 and 7 T for the oxygen annealed (a) $La_{1.4}Sr_{1.6}Mn_2O_7$, (b) $La_{1.4}Sr_{1.2}Ca_{0.4}Mn_2O_7$ and (c) $La_{1.4}Sr_{0.8}Ca_{0.8}Mn_2O_7$ (in this case the 0.1 T $\rho$ *vs. T* curve is also reported). In the insets: correspondent MR curves *vs. T* for the same samples.

**Figure 12** – Molar magnetisation *M/H vs. T* curves at 100 Oe for the oxygen annealed $La_{1.4}Sr_{1.6}Mn_2O_7$ (white squares), $La_{1.4}Sr_{1.2}Ca_{0.4}Mn_2O_7$ (black circles) and $La_{1.4}Sr_{0.8}Ca_{0.8}Mn_2O_7$ (black squares).

**Figure 13** – Molar magnetisation *M/H vs. T* curves at 100 Oe for the argon annealed $La_{1.4}Sr_{1.6}Mn_2O_7$ (white squares), $La_{1.4}Sr_{1.2}Ca_{0.4}Mn_2O_7$ (black circles) and $La_{1.4}Sr_{0.8}Ca_{0.8}Mn_2O_7$ (black squares).

# Table Caption

**Table 1 -** Lattice constants (*a*, *b* and *c*), cell volume (*V*) and *c/a* values for as-prepared (AP), oxygen (O) and argon (Ar) annealed $(La)_{1.4}(Sr_{1-y}Ca_y)_{1.6}Mn_2O_{7\pm\delta}$ (*y*=0, 0.25 and 0.50) samples.

**Table 2 -** Resistivity values at 300 and at 10 K, $T_\rho$ and $T_C$ values, temperature and % values of the MR peaks at 7T and at 1T for the oxygen (O) and argon (Ar) annealed $(La)_{1.4}(Sr_{1-y}Ca_y)_{1.6}Mn_2O_{7\pm\delta}$ (*y*=0, 0.25 and 0.50) samples.

## Table 1

| Sample | *a* (Å) | *b* (Å) | *c* (Å) | *V* (Å$^3$) | *c/a* |
|---|---|---|---|---|---|
| *y*=0     AP | 3.8682(2) | 3.8682(2) | 20.274(1) | 303.3(3) | 5.241(1) |
| *y*=0.25  AP | 3.8640(1) | 3.8640(1) | 20.1366(7) | 300.6(9) | 5.211(3) |
| *y*=0.50  AP | 3.8715(1) | 3.8715(1) | 19.8195(7) | 297.0(2) | 5.119(2) |
| *y*=0     O (62%) | 3.8683(5) | 3.8688(5) | 20.175(3) | 301.7(1) | 5.215(3) |
| *y*=0     O (38%) | 3.8666(6) | 3.8698(7) | 20.016(2) | 299.4(2) | 5.172(4) |
| *y*=0.25  O (60%) | 3.8598(4) | 3.8610(4) | 19.880(2) | 296.2(1) | 5.149(3) |
| *y*=0.25  O (40%) | 3.8601(4) | 3.8618(5) | 20.011(2) | 298.3(1) | 5.182(3) |
| *y*=0.50  O | 3.8573(3) | 3.8556(2) | 19.693(1) | 292.8(1) | 5.105(2) |
| *y*=0     Ar | 3.8635(1) | 3.8635(1) | 20.3576(6) | 303.8(8) | 5.269(1) |
| *y*=0.25  Ar | 3.8642(1) | 3.8640(1) | 20.1953(8) | 301.5(1) | 5.226(1) |
| *y*=0.50  Ar | 3.8696(4) | 3.8696(4) | 19.9810(8) | 299.1(1) | 5.164(3) |

## Table 2

| Sample | $\rho_{300K}$ (Ωcm) | $\rho_{10K}$ (Ωcm) | $T_\rho$ (K) | $MR_{peak7T}$ *T* (K) and (%) | $MR_{peak1T}$ *T* (K) and (%) | $T_C$ (K) |
|---|---|---|---|---|---|---|
| *y*=0     O | 1.64 | 4.71 | 146.2 | - | - | 56 |
| *y*=0.25  O | 0.71 | 1.84 | 141.5 | - | - | 33 |
| *y*=0.50  O | 0.39 | 0.51 | 134.5 | 74 K (62) | 65 K(35) | - |
| *y*=0     Ar | 927 | $3.2\cdot10^6$ | 83 | 98 K(59.7) | 69 K(30.5) | 53.5 |
| *y*=0.25  Ar | 2.2 | $2.5\cdot10^5$ | 71 | 70 K(99.5) | 63 K(81.4) | 30.5 |
| *y*=0.50  Ar | 0.9 | 112 | 100 | 112 K(54.3) | - | 24 |

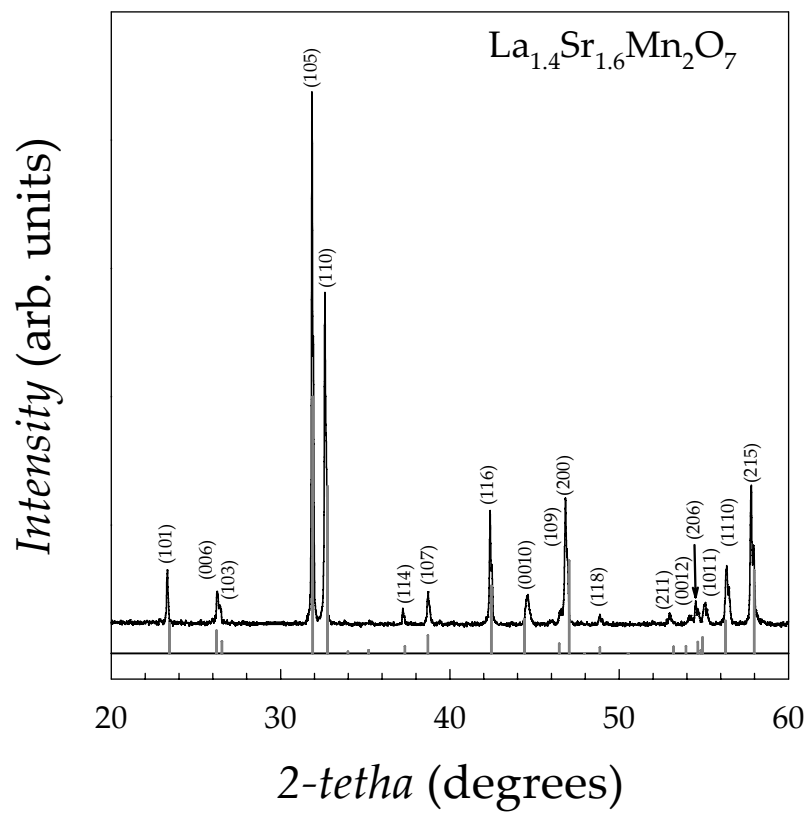

**Figure 1**

**Figure 2**

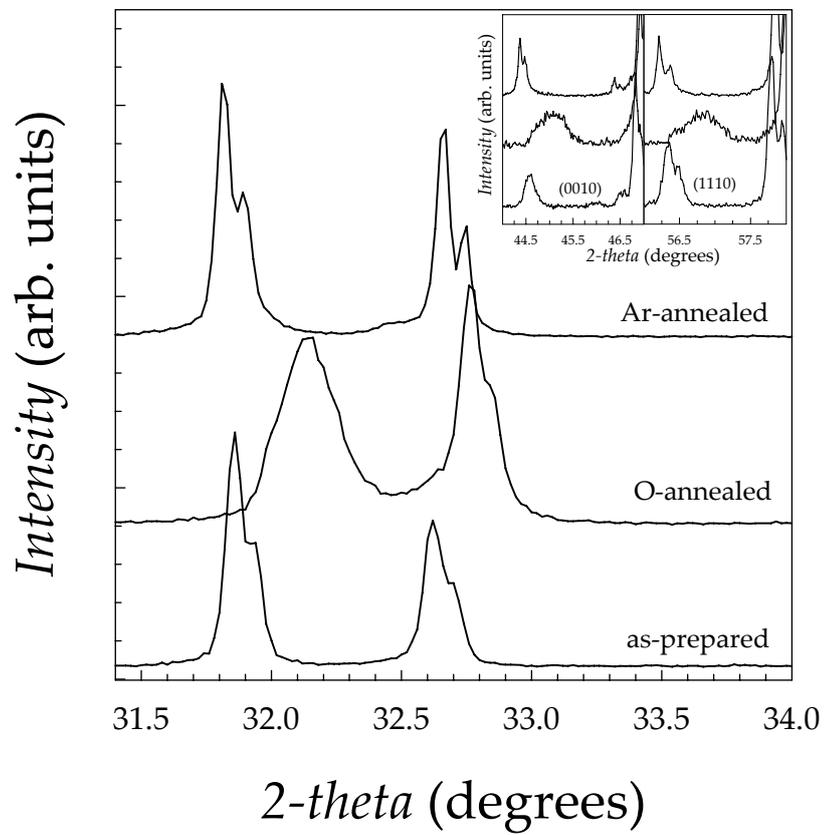

**Figure 3**

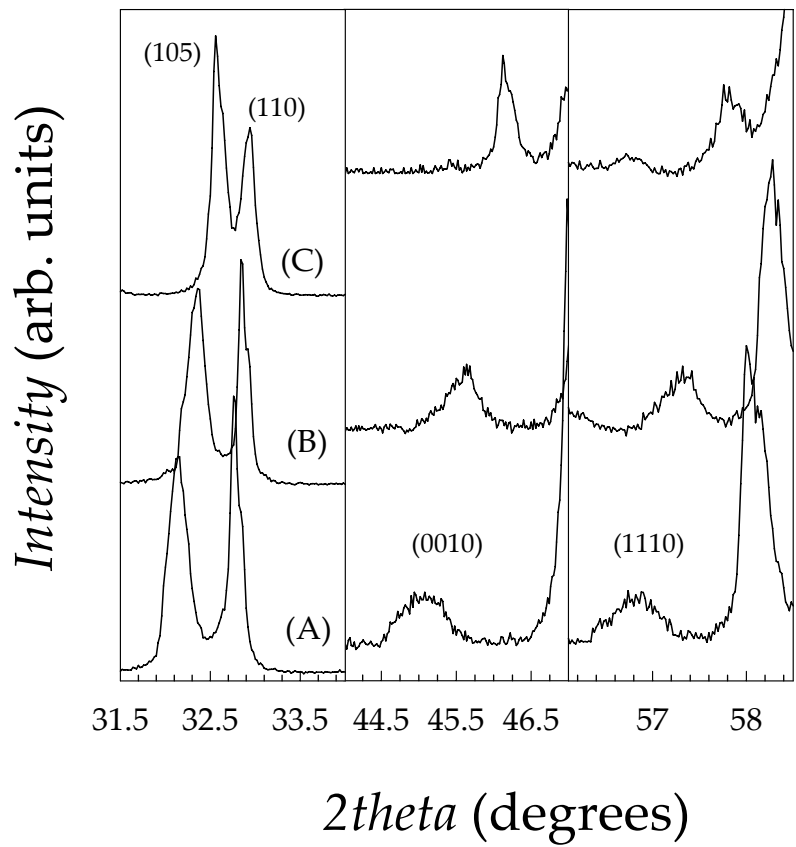

**Figure 4**

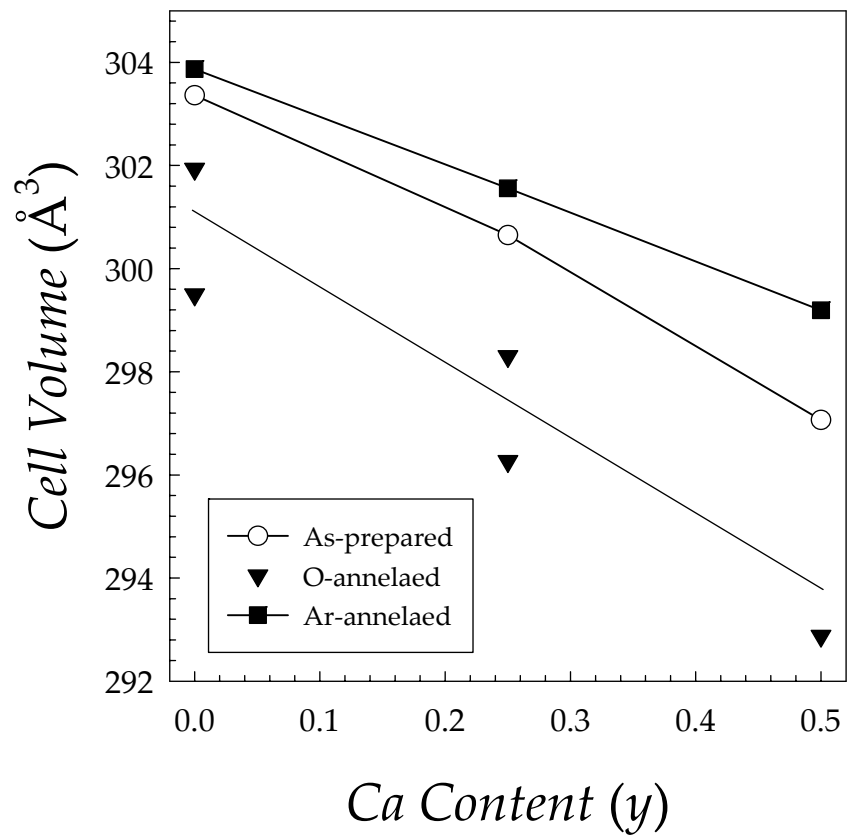

**Figure 5**

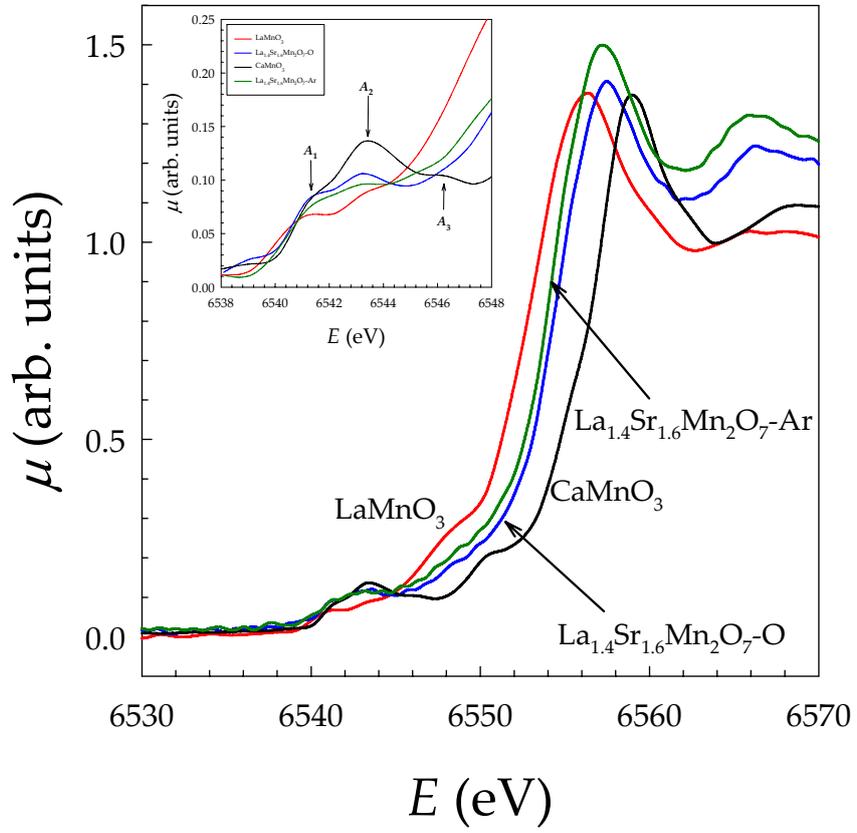

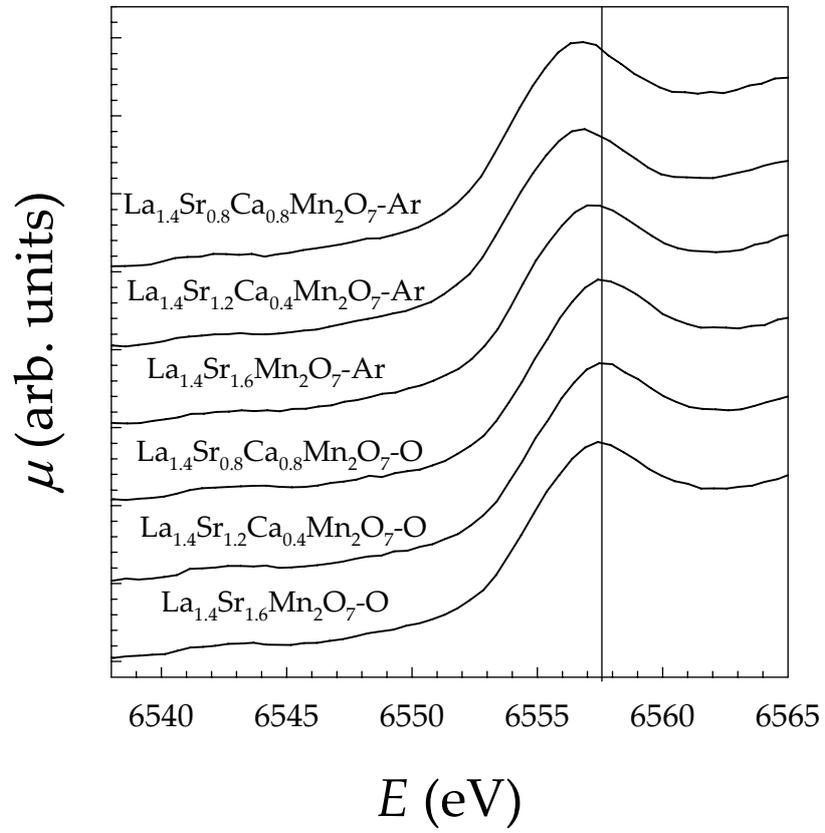

$\mu$ (arb. units)

$La_{1.4}Sr_{0.8}Ca_{0.8}Mn_2O_7$-Ar

$La_{1.4}Sr_{1.2}Ca_{0.4}Mn_2O_7$-Ar

$La_{1.4}Sr_{1.6}Mn_2O_7$-Ar

$La_{1.4}Sr_{0.8}Ca_{0.8}Mn_2O_7$-O

$La_{1.4}Sr_{1.2}Ca_{0.4}Mn_2O_7$-O

$La_{1.4}Sr_{1.6}Mn_2O_7$-O

$E$ (eV)

**Figure 7**

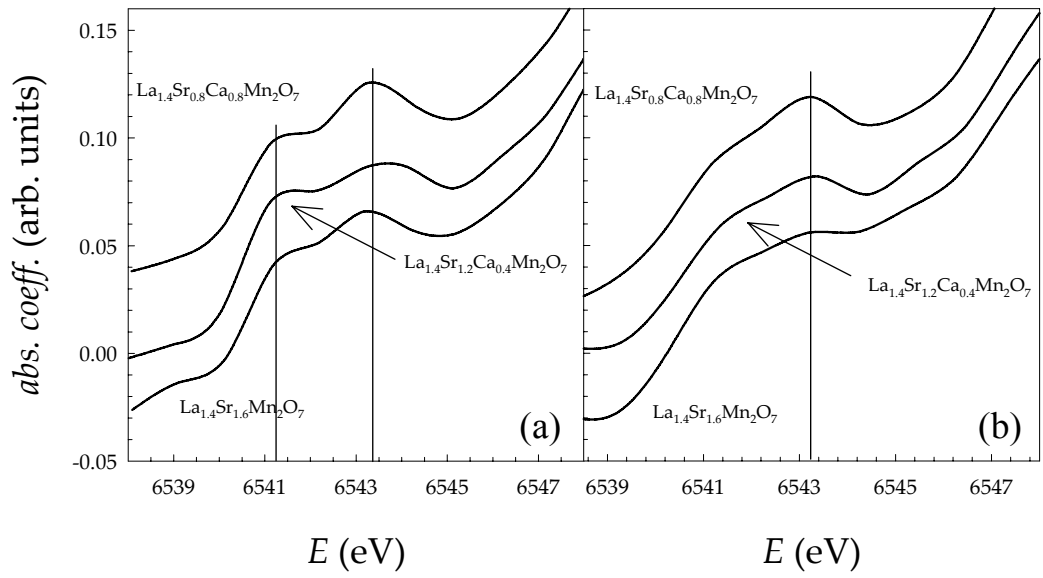

**Figure 8**

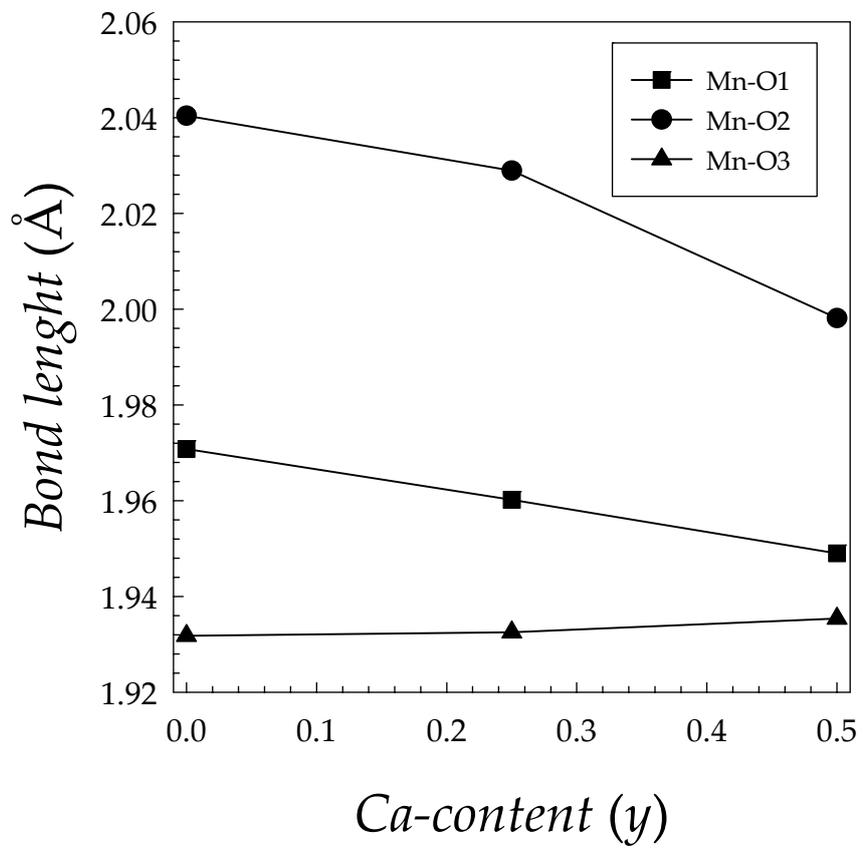

**Figure 9**

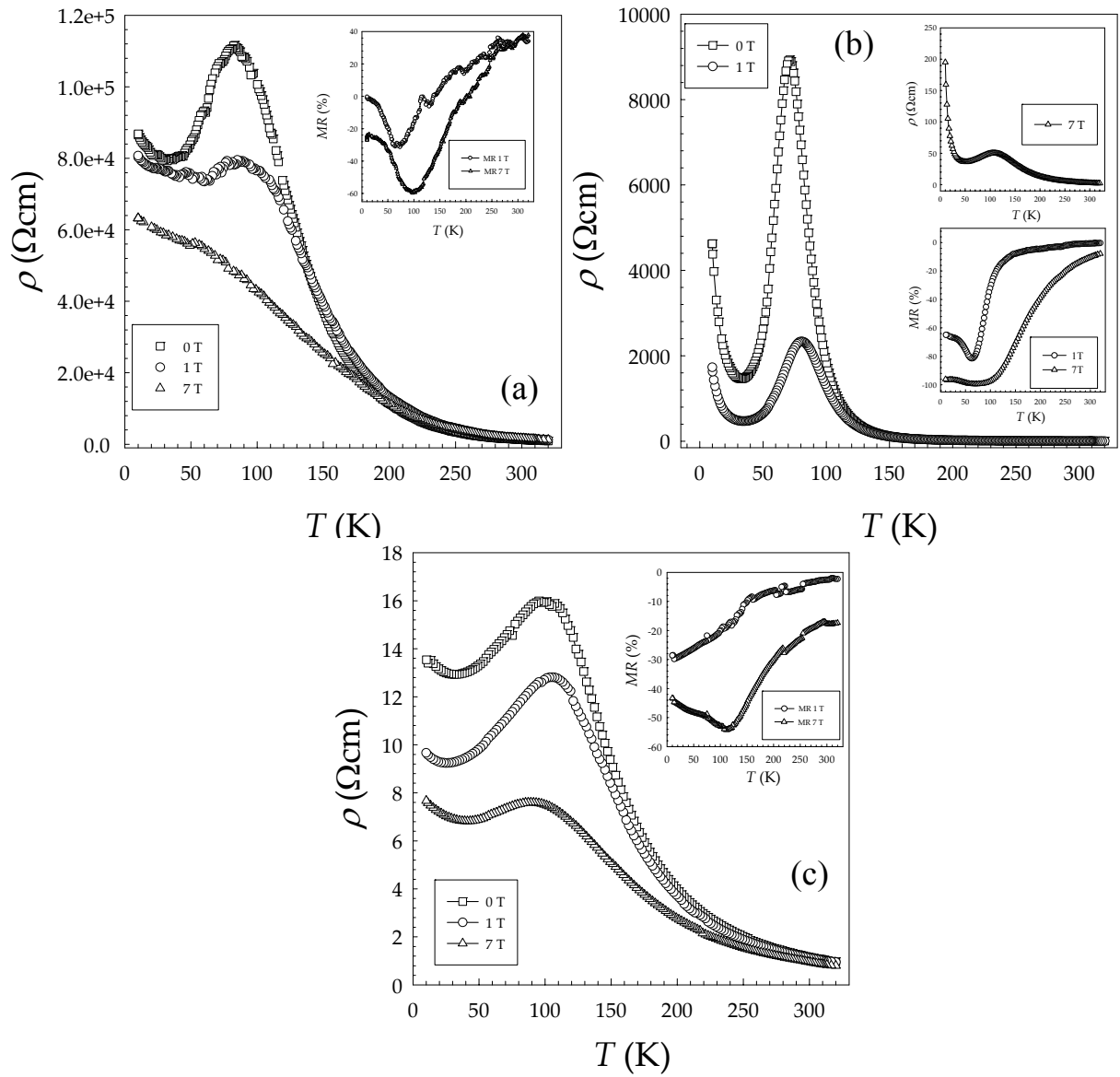

**Figure 10**

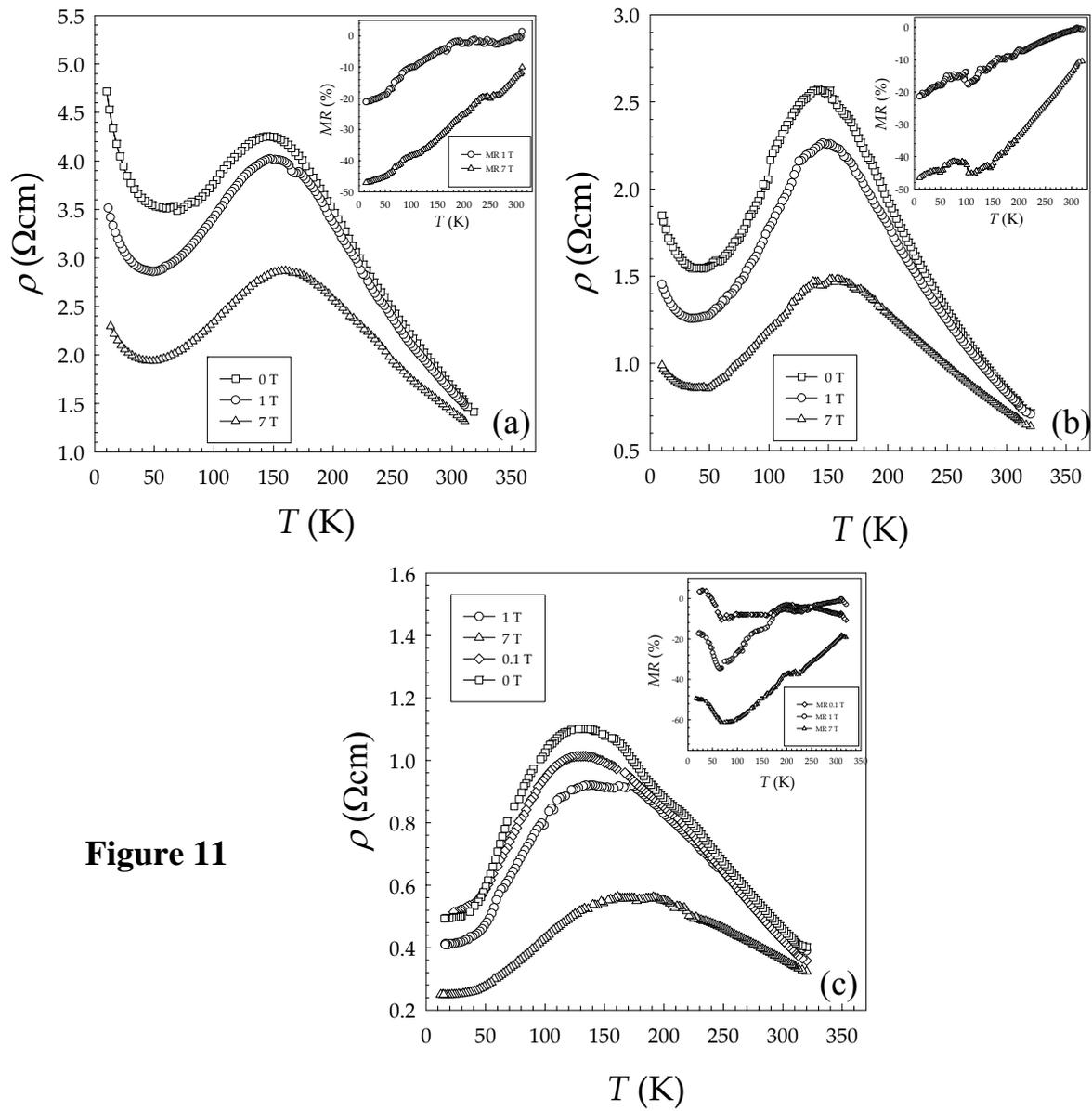

**Figure 11**

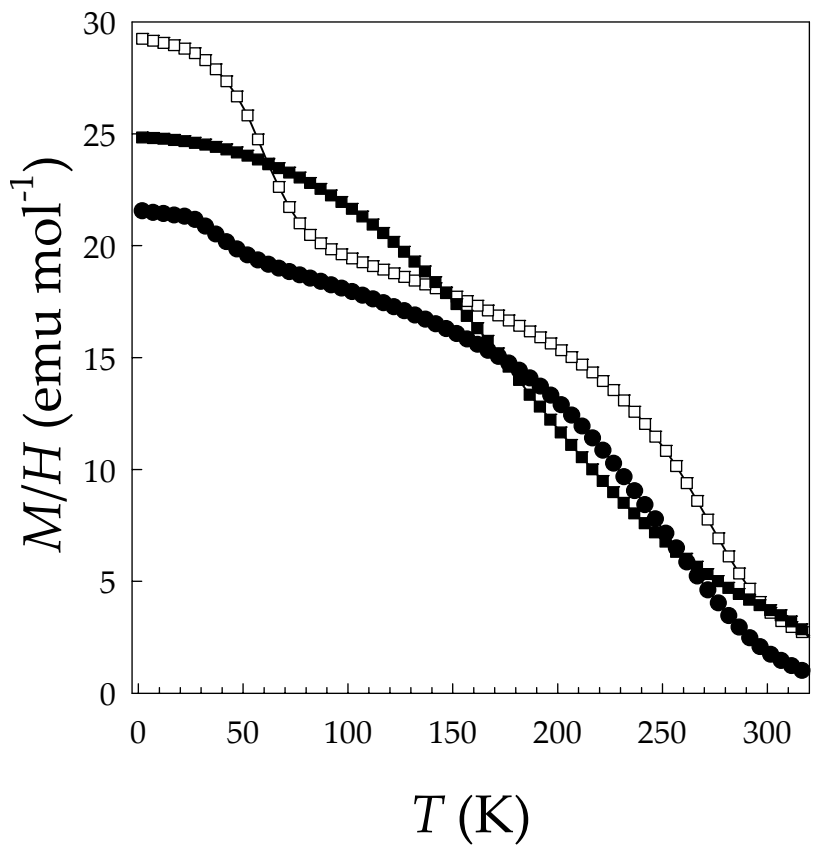

**Figure 12**

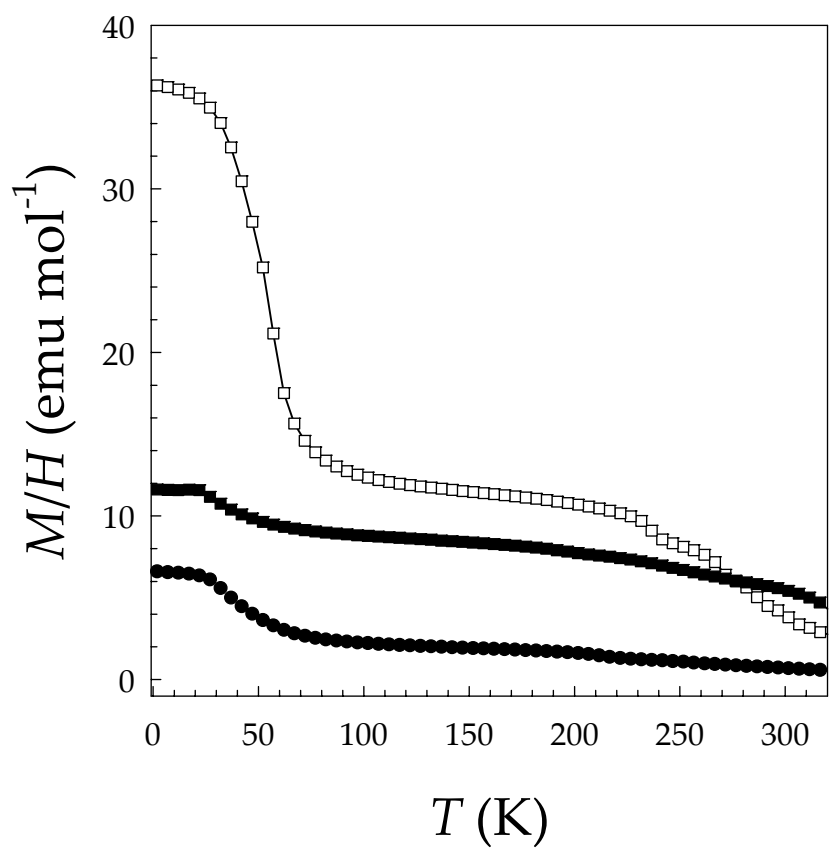

**Figura 13**